%% file: main.tex
\documentclass[conference]{IEEEtran}
\IEEEoverridecommandlockouts
\usepackage{cite}
\usepackage{amsmath,amssymb,amsfonts,mathrsfs}
\usepackage{algorithmic}
\usepackage{algorithm}
\usepackage{array}
\usepackage{graphicx}
\usepackage{textcomp}
\usepackage{tabularx}
\usepackage{xcolor}
\usepackage[caption=false,font=normalsize,labelfont=sf,textfont=sf]{subfig}
\usepackage{stfloats}
\usepackage{siunitx}
\usepackage{url}
\usepackage{verbatim}
\usepackage{amssymb}
\usepackage{bm}
\usepackage{longtable}
\usepackage{threeparttable}
\usepackage{colortbl}
\usepackage{booktabs}
\usepackage{multirow}
\usepackage{mathptmx}
\usepackage[T1]{fontenc}
\usepackage[backref=page]{hyperref}
\usepackage{tabularray}

\hypersetup{
colorlinks,
linkcolor=blue,
citecolor=blue,
anchorcolor=blue,
urlcolor=blue,  
}
\usepackage{orcidlink} % 必须在hyperref之后加载

\def\BibTeX{{\rm B\kern-.05em{\sc i\kern-.025em b}\kern-.08em
    T\kern-.1667em\lower.7ex\hbox{E}\kern-.125emX}}
\begin{document}

\title{HLC: A High-Quality Lightweight Mezzanine Codec Featuring High-Throughput Palette}

%%%%%%%%% AUTHOR - PLEASE UPDATE
\author{\textsuperscript{1}Chenlong~He\,\orcidlink{0009-0002-2721-5709},
        \textsuperscript{2}Leilei~Huang\,$^\star$\,\orcidlink{0000-0002-8900-4109},
        \textsuperscript{1}Wei~Li\,\orcidlink{0000-0002-6861-7117},
        \textsuperscript{1}Hanyang~Cui\,\orcidlink{0009-0008-9767-2087},
        \textsuperscript{3}Zhijian~Hao\,\orcidlink{0000-0002-7892-5973},
        \textsuperscript{1}Xiaoyang~Zeng\,\orcidlink{0000-0003-3986-137X}\,~\IEEEmembership{Member,~IEEE}, \\and 
        \textsuperscript{1}Yibo~Fan\,$^\star$\,\orcidlink{0000-0003-2523-8261}\,~\IEEEmembership{Member,~IEEE}
        \\
        \textsuperscript{1}Fudan University, Shanghai \textsuperscript{2}East China Normal University, Shanghai \textsuperscript{3}Xidian University, Xi'an
        \\
        \textsuperscript{*}Corresponding authors: \texttt{fanyibo@fudan.edu.cn, llhuang@cee.ecnu.edu.cn}
        \thanks{This article has been accepted for publication in the 2026 IEEE International Symposium on Circuits and Systems (ISCAS 2026). \textcopyright~2026 IEEE. Personal use of this material is permitted. Permission from IEEE must be obtained for all other uses, in any current or future media, including reprinting/republishing this material for advertising or promotional purposes, creating new collective works, for resale or redistribution to servers or lists, or reuse of any copyrighted component of this work in other works.}
        % <-this % stops a space
}

\maketitle

%%%%%%%%% CONTENT - PLEASE UPDATE
\input{sec/0_abstract}
\input{sec/1_intro}
\input{sec/2_proposed_method}
\input{sec/3_experiment}
\input{sec/4_conclusion}

\bibliographystyle{IEEEtran}
\bibliography{main}

\end{document}

%% file: sec/0_abstract.tex
\begin{abstract}
\label{abstract}
Existing mezzanine image codecs lack specialized screen content coding tools and therefore struggle to maintain high image quality under bandwidth constraints, especially in areas with dense text. Although distribution codecs offer advanced screen content compression techniques, their high computational complexity makes them impractical for mezzanine coding. To address this shortfall, we introduce the High-quality Lightweight Codec (HLC), a solution centered on enabling practical, high-throughput palette for mezzanine coding. The core innovation is a novel data-dependency-free palette that eliminates the throughput bottlenecks. To ensure its effectiveness across all content, a co-designed rate-distortion optimization module arbitrates between the palette and traditional prediction modes, while a data reuse strategy between rate estimation and entropy coding minimizes the overall hardware resources required for the system. Experimental results show that, compared with a 4K@120fps JPEG-XS encoder, HLC achieves the same throughput while using only half the LUT resources and delivers BD-PSNR improvements of 3.461dB, 3.299dB, and 5.312dB on gaming, natural, and text content datasets, respectively.
\end{abstract}
\begin{IEEEkeywords}
Image Codec, Lightweight Compression, Mezzanine Coding, FPGA
\end{IEEEkeywords}

%% file: sec/1_intro.tex
\section{Introduction}
\label{sec:intro}
\IEEEPARstart{D}{espite} increasing communication bandwidth in the mezzanine workflow, it remains insufficient to meet the rapid growth of video resolution and frame rate. Neither serial digital interface (SDI)-based wired nor internet protocol (IP)-based wireless transmission can handle the bandwidth demands of uncompressed video~\cite{jpeg_xs_std}, posing new challenges for mezzanine coding.\par
In general, the demands of mezzanine coding are fourfold: \textbf{1) High Quality:} achieving 10--30$\times$ compression ratios across diverse content while maintaining high visual quality; \textbf{2) Lightweight Design:} minimizing hardware resource consumption for edge deployment; \textbf{3) High Throughput:} ensuring high data rates for real-time processing; and \textbf{4) Frame Independence:} requiring each frame to be randomly accessed and modified without affecting others. This final requirement for frame-level autonomy is why image codecs are typically used. Due to these multidimensional requirements, although various compression standards~\cite{hevc_std,jpeg_2000_std,plc_std,jpeg_xs_std} and their implementations~\cite{jpeg_xs_hw_fpga_2023,intopix2021,jpeg_2000_hw_fpga_2025,h265_i_p_hw_asic_2024} have been introduced, few fully meet all these demands.\par
According to their application scenarios as defined in~\cite{jpeg_xs_std}, existing codecs are classified into distribution codecs~\cite{hevc_std}, and mezzanine codecs~\cite{jpeg_2000_std,plc_std,jpeg_xs_std}. 
Distribution codecs, which typically employ a block-based hybrid coding architecture, integrate a rich set of complex prediction and entropy coding tools. This enables them to achieve extremely high compression efficiency across most content. However, this architectural complexity inherently conflicts with the core mezzanine requirements of Lightweight Design and High Throughput, rendering them unsuitable for such applications.\par
In contrast, mezzanine codecs like JPEG-XS~\cite{jpeg_xs_std} prioritize low latency and hardware simplicity. They achieve this by adopting a streamlined architecture that dispenses with the concept of blocks and predictive coding entirely. While this approach allows them to meet real-time throughput demands with minimal resources, the absence of advanced coding tools severely compromises their ability to compress screen content efficiently, leading to poor quality, especially in text-rich areas.\par
The unique requirements of mezzanine coding call for the development of a specially designed image codec, driving the design of our high-quality, lightweight solution, HLC. Our work bridges the gap between high-complexity distribution codecs and lightweight mezzanine codecs by merging their respective strengths through targeted innovations.
The primary contributions of our solution are as follows: 
\begin{itemize}
    \item \textbf{A High-Throughput, Data-Dependency-Free Palette Architecture:} To overcome the data dependency bottleneck that limits the throughput of conventional palette architectures, we propose a novel design featuring a virtual cluster table. By eliminating the dependency in the pixel clustering stage, our method enables a fully parallelized hardware implementation.
    \item \textbf{A Co-Designed RDO for Effective Palette Integration:} To ensure the effective integration of our palette, we introduce a co-designed RDO. The co-design involves accurately modeling the rate cost of palette and precisely tuning the QP-$\lambda$ table for an optimal rate-distortion trade-off to leverage the palette's strengths on screen content while preserving high quality on natural images.
\end{itemize}
The effectiveness of these contributions is validated by our experimental results: when compared to a 4K@120fps JPEG-XS encoder~\cite{jpeg_xs_hw_fpga_2023}, HLC matches its throughput with only half the LUT resources, while delivering significant BD-PSNR improvements of 3.461dB, 3.299dB, and 5.312dB on gaming, natural, and text content, respectively.

%% file: sec/2_proposed_method.tex
\section{Architectural and Algorithmic Co-Design}
\label{sec:prop_method}
\begin{figure*}[t!]
\centering
\includegraphics[width=0.9\textwidth]{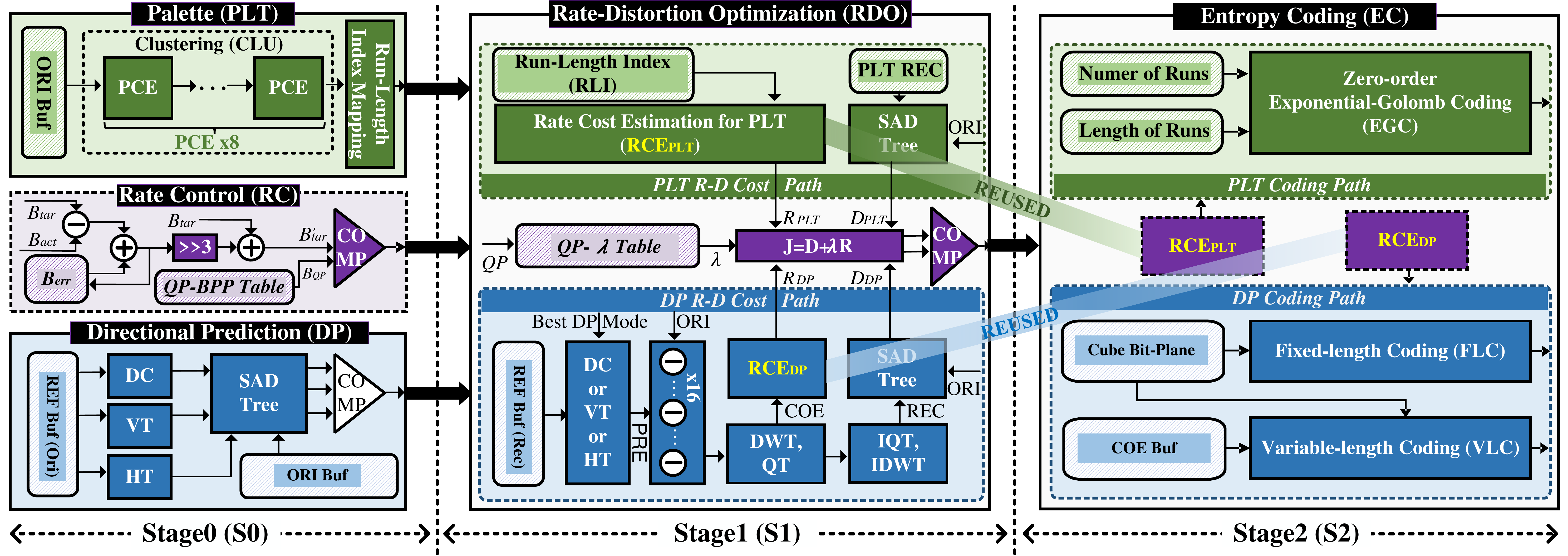}
\vspace{-10pt}
\caption{Hardware architecture of HLC, which contains three pipeline stages.}
\vspace{-10pt}
\label{fig:hw_arh}
\end{figure*}
\begin{figure}[t!]
\centering
\includegraphics[width=\linewidth]{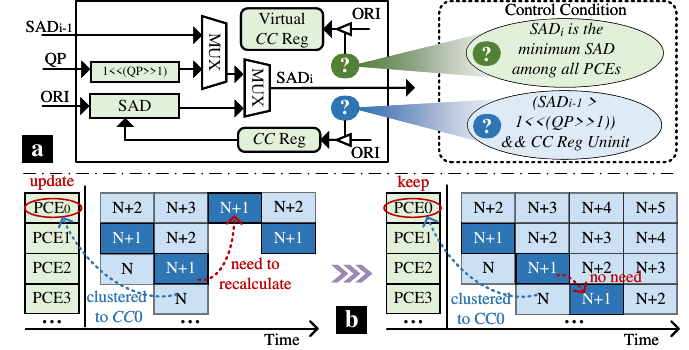}
\vspace{-20pt}
\caption{Hardware architecture and pipeline space-time diagram of pixel clustering engine (PCE). (a) Architecture. (b) Space-time diagram.}
\vspace{-10pt}
\label{fig:pce_arc}
\end{figure}
HLC employs a hybrid coding architecture, as illustrated in Fig.~\ref{fig:hw_arh}. The architecture is organized into a three-stage pipeline, corresponding to stages \textbf{S0}, \textbf{S1}, and \textbf{S2}. This framework integrates a suite of coding tools, including palette (PLT), rate control (RC), directional prediction (DP), rate-distortion optimization (RDO), and entropy coding (EC). The coding unit (CU) for the pipeline is a 16$\times$4 block. This design is deliberate: while it processes the same number of pixels as a 8$\times$8 block found in standards like HEVC~\cite{hevc_std}, its shape halves the required line buffer depth from eight lines to four.
\begin{itemize}
    \item In \textbf{S0}, PLT and DP are employed to eliminate spatial-domain redundancy, while RC is designed to precisely control the bitrate.
    \item In \textbf{S1}, RDO is adopted to achieve the optimal combination of coding tools. Moreover, multidimensional discrete wavelet transform (DWT) and quantization (QT) are employed to further reduce frequency-domain redundancy.
    \item In \textbf{S2}, EC is specifically designed to maximally remove entropy redundancy, while remaining highly parallel.
\end{itemize}
\subsection{High-Throughput PLT via a Dependency-Free Architecture}
The palette (PLT) is a tool first introduced in distribution standards like HEVC~\cite{hevc_std} to efficiently handle screen content. For a given block of pixels, PLT creates a small table of representative colors (clusters) and then represents each pixel by an index into that table, effectively reducing spatial redundancy. However, the core process of pixel clustering introduces data dependencies that are a major obstacle for high-throughput hardware implementation. The only known prior design~\cite{plt_hardware} achieves just 1080P@30fps while consuming 66K LUTs, making conventional PLT unsuitable for mezzanine coding.\par
To solve this, we propose a data-dependency-free PLT architecture. It features a novel pixel clustering strategy and a dedicated hardware engine that achieves high throughput at a low hardware cost. Our final implementation delivers a 3.983dB BD-PSNR gain on text content and processes 4K@120fps on a KC705 FPGA, using only 15.4K LUTs.\par
\subsubsection{The Data Dependency Bottleneck in Pixel Clustering}
The  hardware unit of our PLT module is the Pixel Clustering Engine (PCE), whose architecture is shown in Fig.~\ref{fig:pce_arc}(a). The PCE's task is to group the pixels of a CU into a maximum of eight clusters, each defined by a Cluster Center (CC). The process is as follows: each pixel is compared against all existing CCs using the Sum of Absolute Differences (SAD). If the SAD is below a QP-derived threshold---defined as $(1 \ll (QP \gg 1))$---the pixel is assigned to the cluster with the minimum SAD. If the pixel cannot be assigned and fewer than eight CCs exist, the pixel itself establishes a new CC. If a ninth CC is required, the CU is deemed unsuitable for PLT.\par
In a conventional PLT algorithm, the CC value is updated whenever a new pixel is assigned to its cluster. This creates a critical data dependency that stalls parallel processing. As illustrated in the left half of Fig.~\ref{fig:pce_arc}(b), an update to a CC by one PCE forces subsequent pixels being processed by other PCEs to restart their calculations with the new CC value, causing a pipeline flush and destroying throughput.\par
\subsubsection{A Virtual Cluster Table for Dependency-Free Clustering}
To break this dependency, we introduce a \textbf{virtual cluster table} strategy. As shown in Fig.~\ref{fig:hw_arh}(a), each PCE maintains two registers: a standard \texttt{CC Reg} and a \texttt{Virtual CC Reg}. The standard \texttt{CC Reg} is written to only once when a new cluster is created; its value remains static for all clustering decisions within the CU. The \texttt{Virtual CC Reg} is updated continuously but is used only for pixel reconstruction at the end, not for the clustering decisions themselves. Since the CC value used for SAD calculations is fixed, the data dependency is completely eliminated. This enables a fully pipelined, high-throughput design where no recalculation is needed.\par
This dependency removal introduces a minor, acceptable trade-off in compression efficiency. Because clustering decisions are based on the initial CC value rather than a continuously updated average, there is a slight performance drop. Under the evaluation setup in Section~\ref{sec:setup}, we measure this loss to be 0.121dB in BD-PSNR on the text dataset.\par
\subsubsection{Compressing Cluster Indices via Run-Length Mapping}
\begin{figure}[t!]
\centering
\includegraphics[width=\linewidth]{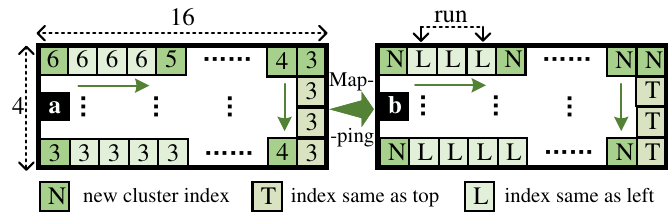}
\vspace{-20pt}
\caption{An illustration of run-length index mapping. (a) Pixels of CU represented by corresponding cluster index. (b) Mapping results. }
\vspace{-10pt}
\label{fig:run_len_map}
\end{figure}

Once the cluster table for a CU is finalized, each pixel is represented by its corresponding cluster index, as shown in Fig.~\ref{fig:run_len_map}(a). To further compress the entropy redundancy present in this 2D map of indices, we propose a run-length index mapping scheme. Each pixel's index is mapped to one of three symbols: `0` (L) if it matches the left neighbor, `1` (T) for a match with the top neighbor, or `2` (N) otherwise. As illustrated in Fig.~\ref{fig:run_len_map}(b), this mapping effectively converts horizontal spatial redundancy into runs of identical symbols, which are then compressed efficiently during rate cost estimation and entropy coding.\par
\subsection{Effective PLT Integration via a Co-Designed RDO}
To enable the PLT module for screen content without degrading performance on natural images, we introduce a Rate-Distortion Optimization (RDO) framework. This framework intelligently selects the optimal coding mode for each CU. The strategy is so effective that it results in a 0.345dB BD-PSNR gain even on natural content for PLT, as shown in Table~\ref{tab:hardware_algorithm_comparison}. The success of this RDO hinges on two components: an accurate Rate Cost Estimation (RCE) and a well-tuned QP-$\lambda$ relationship. Furthermore, we mitigate the high hardware cost typical of RDO by using a hardware-friendly DWT (enabled by our 16$\times$4 CU design) and by reusing RCE results in the entropy coding, as shown in  Fig.~\ref{fig:hw_arh}.
\subsubsection{Rate-Distortion Cost Estimation}
The RDO process selects the optimal mode by comparing the rate-distortion cost for both DP and PLT. The Distortion cost ($D$) is measured as the SAD between the original and reconstructed pixels. The Rate cost $R$ is estimated by two separate modules: RCE\textsubscript{DP} and RCE\textsubscript{PLT}. Both RCE modules are designed to pass intermediate results directly to EC to reduces hardware resources.\par
The specific estimation processes are as follows:
\begin{itemize}
    \item \textbf{For RCE\textsubscript{DP}}, the CU is divided into sixteen 2$\times$2 coefficient cubes. The rate cost ($R_{DP}$) is calculated as the sum of the bit-widths of all coefficients. The maximum bit-width within each cube is recorded as a bit-plane value, which is the intermediate result reused by EC.
    \item \textbf{For RCE\textsubscript{PLT}}, the process operates on the run-length index (RLI) map illustrated in Fig.~\ref{fig:run_len_map}. The rate cost ($R_{PLT}$) is the sum of the bit-widths required to represent the length of each run of identical indices. Both the total number of runs and their lengths are reused by EC.
\end{itemize}
\subsubsection{\texorpdfstring{$QP-\lambda$}{QP-lambda} Table Tuning}
\begin{figure}[t!]
\centering
\includegraphics[width=\linewidth]{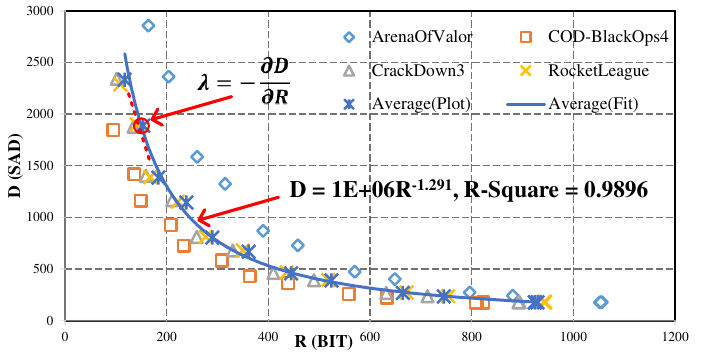}
\vspace{-20pt}
\caption{Fitting results of the $R$-$D$ curve and how to get $QP-\lambda$ table.}
\label{fig:qp_lambda_tune}
\vspace{-10pt}
\end{figure}

The accuracy of the RDO depends on the Lagrange multiplier ($\lambda$), which balances the R-D trade-off and is equal to the slope of the R-D curve~\cite{rc_rlmd_1}. To establish a robust $QP$-$\lambda$ relationship for HLC, we model the average R-D curve from a training set of diverse images. As shown in Fig.~\ref{fig:qp_lambda_tune}, the data closely fits the model $D = C R^{-K}$ (R-Square = 0.9896). Based on this model, we pre-calculate an optimized $QP$-$\lambda$ table by deriving the slope at each point on the curve, ensuring an effective balance for the RDO decision.\par
\subsection{Hardware Cost Reduction via RDO-to-EC Data Reuse}
The Entropy Coding (EC) stage further compresses data from RDO. While variable-length coding in distribution codecs often creates hardware and throughput bottlenecks, our EC architecture avoids this. The core of our design is a hybrid fixed- and variable-length coding strategy. For both DP and PLT data streams, we first transmit fixed-length syntax elements that define the length of the variable-length data to follow, enabling parallel processing.
A key innovation is that these fixed-length statistics are directly reused from the RCE engines. This data reuse strategy saves significant hardware resources and breaks the traditional EC throughput bottleneck.
\begin{itemize}
    \item \textbf{For DP data}, the reused bit-plane for each 2x2 cube signals the length of the variable-length coefficients.
    \item \textbf{For PLT data}, the reused run counts and lengths are encoded using zero-order Exponential-Golomb Coding (EGC) to structure the bitstream.
\end{itemize}
This strategy reduces the EC module's hardware cost by an estimated 10.2K LUTs compared to a non-reusing design, with our final implementation consuming only 17.4K LUTs.\par
\subsection{Miscellaneous}
To reduce hardware cost, several simplifications are implemented: adopting three core modes from distribution codecs~\cite{hevc_std} for DP, namely direct current (DC), vertical (VT), and horizontal (HT); and employing the RC from~\cite{h265_i_p_hw_asic_2024}, which adjusts QP of each CU based on the target bits per pixel ($B_{\text{tar}}$) and the accumulated bit error ($B_{\text{err}}$).

%% file: sec/3_experiment.tex
\section{Experiments}
\label{sec:experiments}
\subsection{Experiment Setup}
\label{sec:setup}
\begin{table*}[t!]
  \centering
  \caption{Hardware and Algorithm Performance Comparison}
    \begin{threeparttable}
    \begin{tabular}{cccccccccc}
    \toprule
    \toprule
    \multicolumn{2}{c}{Category} & \multicolumn{7}{c}{Mezzanine Codec} & Distribution Codec \\
    \midrule
    \multicolumn{2}{c}{Standard} & \multicolumn{2}{c}{HLC\textsuperscript{1} (proposed)} & \multicolumn{2}{c}{HLC(w/o PLT)} & JPEG-XS~\cite{jpeg_xs_hw_fpga_2023} & JPEG-2000\textsuperscript{2}~\cite{intopix2021,jpeg_xs_std} & JPEG-2000~\cite{jpeg_2000_hw_fpga_2025} & HEVC-Intra\textsuperscript{3}~\cite{h265_i_p_hw_asic_2024} \\
    \midrule
    \multicolumn{2}{c}{Codec} & Encoder & Decoder & Encoder & Decoder & Encoder & Encoder & Encoder & Encoder \\
    \midrule
    \multicolumn{2}{c}{Platform} & KC705 & KC705 & KC705 & KC705 & Alveo U50 & Arria V 5AGXA7 & KC705 &  KC705 \\
    \midrule
    \multicolumn{2}{c}{Throughput} & 4K@120 & 4K@120 & 4K@120 & 4K@120 & 4K@120 & 4K@60 & 512x512@53 & 4K@2 \\
    \midrule
    \multicolumn{2}{c}{Frequency} & 300MHz & 300MHz & 300MHz & 300MHz & 196MHz & 200MHz & 120MHz  & 69MHz \\
    \midrule
    \multicolumn{2}{c}{Memory} & 9388Kb & 4256Kb & 8802Kb & 3990Kb & 15952Kb & 12977Kb & 5569Kb & 4691Kb \\
    \midrule
    \multirow{3}[6]{*}{Resource} & LUT   & 82K   & 50K & 58K & 43K & 172K & 174K & 79K  & 108K \\
\cmidrule{2-10}
          & REG   & 108K   & 46K & 76K & 35K & 85K & - & 49K & 53K \\
\cmidrule{2-10}
          & DSP   &  6     & 2   & 0 & 2 & 43 & - & - & 798 \\
    \midrule
    \multirow{3}[3]{*}{BD-PSNR$\uparrow$} & TEC~\cite{jvet_com_test_con_scc} & \multicolumn{2}{c}{/} & \multicolumn{2}{c}{-3.983dB} & -5.312dB & -1.766dB & -1.731dB & -0.324dB \\
\cmidrule{2-10}
          & GAC~\cite{barman2021user} & \multicolumn{2}{c}{/} & \multicolumn{2}{c}{-0.813dB} & -3.461dB & 0.194dB & 0.391dB & 6.350dB \\
\cmidrule{2-10}
          & NAC~\cite{jvet_com_test_con_hdr} & \multicolumn{2}{c}{/} & \multicolumn{2}{c}{-0.345dB} & -3.299dB & -0.033dB & 1.941dB & 5.940dB \\
    \bottomrule
    \bottomrule
    \end{tabular}
  \label{tab:hardware_algorithm_comparison}
  \begin{tablenotes}
    \item 1 is chosen as the reference for calculating BD-PSNR in the algorithm performance comparison; 2 is an FPGA-based JPEG-2000 encoder~\cite{intopix2021}, with its datasheet sourced from Section VIII.B of~\cite{jpeg_xs_std}; 3 is a simplified, intra-only HEVC encoder that we reimplemented according to \cite{h265_i_p_hw_asic_2024} and deployed on KC705.
  \end{tablenotes}
  \end{threeparttable}
\vspace{-10pt}
\end{table*}
We conduct a comprehensive evaluation of the proposed HLC against several state-of-the-art hardware encoders. The benchmarks include representative mezzanine codecs (JPEG-2000~\cite{jpeg_2000_hw_fpga_2025,intopix2021} and JPEG-XS~\cite{jpeg_xs_hw_fpga_2023}) and a simplified HEVC intra-only encoder~\cite{h265_i_p_hw_asic_2024} (HEVC-Intra) to represent distribution codecs. The evaluation is twofold. For the hardware analysis, we implement HLC on a KC705 FPGA to measure its throughput and resource utilization. For the algorithm analysis, we measure compression efficiency using the BD-PSNR metric~\cite{gisle2001cal} across three distinct datasets for text (TEC)~\cite{jvet_com_test_con_scc}, gaming (GAC)~\cite{barman2021user}, and natural (NAC)~\cite{jvet_com_test_con_hdr} content at four target bitrates with target bits per pixel (BPP) set to 1.75, 1.50, 1.25, and 1.00. HLC serves as the reference baseline, and the complete results are presented in Table~\ref{tab:hardware_algorithm_comparison}.\par
\vspace{-5pt}
\subsection{Comparison and Discussion}
\label{sec:compare}
\subsubsection{Comparison with mezzanine codec} 
\label{sec:hardware_compare}
For text content (TEC) dataset, HLC has a clear advantage in balancing throughput, resource utilization and compression effciency. To be more specific, compared to~\cite{jpeg_xs_hw_fpga_2023}, HLC delivers the same throughput with half the LUT count and achieves BD-PSNR improvement of 5.312dB. Against the 4K@60fps JPEG-2000~\cite{intopix2021}, HLC again halves LUT usage, doubles throughput, and yields 1.766dB BD-PSNR gain. When measured against~\cite{jpeg_2000_hw_fpga_2025}, HLC maintains comparable LUT consumption while substantially increasing throughput and improving BD-PSNR by 1.731dB. For comprehensive comparison, we conduct experiments on gaming content (GAC) and natural content (NAC) datasets. The results indicate that only~\cite{jpeg_2000_hw_fpga_2025} outperforms our method on the NAC dataset by 1.941dB BS-PSNR. However, HLC delivers much higher throughput than it. Moreover, the storage resource in HLC is primarily consumed by line buffers for input pixels and rotating buffers for intermediate data across CU-level pipeline stages. Only the line buffers scale with image width, enabling HLC to support higher-resolution encoding with minimal additional storage.\par
\subsubsection{Comparison with distribution codec}
\label{sec:algorithm_compare}

To evaluate the viability of distribution codecs, we implemented a simplified, intra-only version of a state-of-the-art HEVC encoder~\cite{h265_i_p_hw_asic_2024} on KC705. While this HEVC-Intra design uses a comparable 108K LUTs, its architectural complexity, particularly in the entropy coding stage, creates a severe throughput bottleneck, limiting its performance to just 4K@2fps. This result demonstrates that despite offering superior compression quality on some content, the inherent complexity of distribution codecs makes them fundamentally unsuitable for the real-time demands of mezzanine coding..\par
\subsubsection{Ablation Study}
\label{sec:ablation_study}
As shown in Table~\ref{tab:hardware_algorithm_comparison}, removing the PLT functionality (HLC w/o PLT) causes a significant \textbf{3.983dB} drop in BD-PSNR on text content. This substantial gain confirms that the system-level hardware investment of 24K LUTs is a highly effective trade-off.\par

%% file: sec/4_conclusion.tex
\vspace{-5pt}
\section{Conclusion}
\label{sec:conclusion}
This paper introduces the High-quality, Lightweight Codec (HLC), a hybrid image codec designed to resolve the critical trade-off between compression efficiency, hardware cost, and throughput in mezzanine coding. HLC's architecture strategically integrates a novel \textit{data-dependency-free} PLT, which eliminates the performance bottlenecks of traditional screen content coding. This is complemented by a co-designed RDO and a data reuse strategy that feeds rate estimation results directly to EC. Experimental results validate our approach: HLC achieves the same 4K@120fps throughput as a state-of-the-art JPEG-XS encoder but with only half the LUT resources, while delivering a substantial 5.312dB BD-PSNR gain on text content, providing a superior, balanced solution for modern mezzanine workflows.\par